# Isotopic yield in cold binary fission of even-even $^{244-258}$Cf isotopes

K. P. Santhosh*, Annu Cyriac and Sreejith Krishnan
*School of Pure and Applied Physics, Kannur University, Swami Anandatheertha Campus, Payyanur 670327, Kerala, India*

**Abstract.**
   The cold binary fission of even-even $^{244-258}$Cf isotopes has been studied by taking the interacting barrier as the sum of Coulomb and proximity potential. The favorable fragment combinations are obtained from the cold valley plot (plot of driving potential vs. mass number of fragments) and by calculating the yield for charge minimized fragments. It is found that highest yield for $^{244,246,248}$Cf isotopes are for the fragments with isotope of Pb (Z=82) as one fragment, whereas for $^{250}$Cf and $^{252}$Cf isotopes the highest yield is for the fragments with isotope of Hg (Z=80) as one fragment. In the case of $^{254,256,258}$Cf isotopes the highest yield is for the fragments with Sn (Z=50) as one fragment. Thus, the fragment combinations with maximum yield reveal the role of doubly magic and near doubly magic nuclei in binary fission. It is found that asymmetric splitting is favoured for Cf isotopes with mass number A $\leq$ 250 and symmetric splitting is favoured for Cf isotopes with A > 252. In the case of Cf isotope with A=252, there is an equal probability for asymmetric and symmetric splitting. The individual yields obtained for the cold fission of $^{252}$Cf isotope are compared with the experimental data taken from the γ- γ- γ coincidences technique using Gammasphere.

## 1. Introduction

   More than seventy-five years of research on nuclear fission have clearly shown that, the low energy fission of heavy elements (Z>90) was one of the most complex phenomena of nuclear reactions. Most of the nuclear reactions take place through the binary fission process, a low energy fission, where the fissioning nucleus ends up in two fission fragments and the fragments were formed after the fission barrier has been overcomed. In 1939 Hahn et al., [1] discovered that the uranium atom was fragmented into two parts, which are more or less equal in size. Bohr and Wheeler [2] developed a theory of fission based on the liquid drop model. The authors gave a theory of the effect based on the usual ideas of penetration of potential barriers.

   Experimental studies of cold fission started in the early 80's by Signarbieux et al., [3] Armbruster et al., [4] and found that the relative yields of different fragmentation modes are governed by the available phase space of the system at scission, determined by the nuclear structure properties of the fragments. The cold spontaneous fission of many actinide nuclei into fragments with masses from 70 to 160 were observed and studied [5-9] and found that in these cold decays both the final fragments were in the ground states and confirmed the theoretical predictions by Sandulescu et al., [10,11]. The first direct observation of cold fragmentation in the spontaneous fission of $^{252}$Cf was carried out [7, 8] using the multiple Ge-detector Compact Ball facility at Oak Ridge National Laboratory where four pairs of neutronless fragmentations that of $^{104}$Zr-$^{148}$Ce, $^{104}$Mo-$^{148}$Ba, $^{106}$Mo-$^{146}$Ba and $^{108}$Mo-$^{144}$Ba were observed. Further in 1996 Sandulescu et al., [12] and Dardenne et al., [13] observed cold fragmentation in the spontaneous fission of $^{252}$Cf with the Gammasphere consisting of 72 detectors where the correlations between the two fragments was observed clearly. Sandulescu et al., [12] using a simple cluster model predicted correctly the most important cold fragmentations observed in the spontaneous cold fission of the nucleus $^{252}$Cf, where the double-folding potential barrier with the M3Y nucleon-nucleon forces gave the relative isotopic yields. The results were in good agreement with the experimental data [12,14].

Ramayya et al., [15] observed and presented the evidence for cold binary and ternary fission in the spontaneous fission of $^{252}$Cf using triple gamma coincidence technique with Gammasphere and identified several correlated pairs whose yields were extracted. Gonnenwein et al., [16] observed the presence of doubly magic $^{132}$Sn fragment in the cold fission of $^{252}$Cf, which was predicted some years ago by Kumar et al., [17].

Moller et al., [16,18] reported spontaneous decay of $^{252}$Cf using a twin ionization chamber where two distinct mass regions of cold fission were observed: the first region includes the mass split 96/156 up to 114/138 and second one comprises only a narrow mass range around the mass split 120/132. Mirea et al., [19] computed the cold fission path in the potential energy surface of $^{252}$Cf by using the two-center shell model, based on the idea of the cold rearrangements of nucleons during the cold fission process and obtained a satisfactory agreement with experimental yields by considering variable mass and charge asymmetry beyond the first barrier of the potential surface. Mirea et al., [19] analyzed the data obtained by Hambsch et al., [5] from the cold fission yields of $^{252}$Cf, and showed that the cold fission of $^{252}$Cf is strongly connected with the cold valley of the doubly magic isotope $^{132}$Sn.

The ground state decay properties (nuclear mass, deformation, $\alpha$ decay energy, $\alpha$ decay half-life, spontaneous fission half life etc.) of even-even isotopes of superheavy (SH) elements with Z= 104-170 has been studied by Smolanczuk [20] based on the macroscopic-microscopic model in which a multi dimensional deformation space describing axially symmetric nuclear shapes are used. Within the Hartree-Fock-Bogoliubov (HFB) approach with the finite-range and density-dependent Gogny force with the D1S parameter, a systematic study of 160 heavy and superheavy nuclei was performed by Warda et al.,[21] and the relevant properties of the ground state such as fission barrier, $\alpha$ decay energy, fission and $\alpha$ half lives were discussed. Staszczak et al.,[22] carried out self-consistent Skyrme-HFB calculations to predict main decay modes of even-even superheavy nuclei with $108 \leq Z \leq 126$ and $148 \leq N \leq 188$, to assess their lifetimes and estimated the center of enhanced stability in the superheavy region, thereby predicted the reflection-symmetric mode and the reflection-asymmetric mode as two spontaneous fission modes in superheavy nuclei. Poenaru et al [23] improved the accuracy of alpha and cluster decay half-life of superheavy element with Z >121 by using a semi-empirical formula for $\alpha$ decay and changing the parameters of analytical super asymmetric fission and of the universal curve for cluster decay. The authors improved the spontaneous fission half lives by using nuclear dynamics based on potential barriers computed by the macroscopic–microscopic method and employing various nuclear inertia variation laws. Poenaru et al.,[24,25] analyzed a way to improve the accuracy of evaluated spontaneous fission of nuclei in superheavy region by using the action integral based on cranking inertia and a potential barrier computed within the two-center shell model.

In this manuscript, our work aims to study the isotopic yield in binary fission of even-even $^{244-258}$Cf isotopes by taking the interacting barrier as the sum of Coulomb and proximity potential. The Coulomb and proximity potential has been used extensively for the studies in the areas of alpha decay [26-29], cluster decay [30-34] and spontaneous fission of heavy and superheavy nuclei [35-39]. The calculations in our work was done for Californium (Cf) nuclei which offer interesting possibilities for decay studies due to the closed shell effects of the daughter nucleus ($^{48}$Ca, $^{208}$Pb, $^{132}$Sn) that has been observed [40,41] and predicted [42-44]. The cold binary fission in the region of Californium isotopes is a rare phenomenon where only few measurements are available. The mass asymmetric yield of fission fragments was explained by the magic or doubly magic character of the heavy fragment $^{132}$Sn or one of its neighboring nuclei. On the other hand the usual (not cold) mass asymmetric fission of transuranium isotopes [45-47] has a very special property which was not entirely reproduced until now by any theoretical work. The mass of the heavier fraction of daughter products centers around an A of 136-144 regardless of the mass of the nuclei undergoing fission.

The formalism used for our calculation is described in Sec.2. The results and discussion on the binary fragmentation of even-even $^{244-258}$Cf isotopes are given in Sec.3 and we summarize the entire work in Sec.4.

## 2. The Model

The binary fission is energetically possible only if $Q$ value of the reaction is positive. ie.

$$Q = M - \sum_{i=1}^{2} m_i > 0 \qquad (1)$$

Here $M$ is the mass excess of the parent, $m_i$ is the mass excess of the fragments. The interacting potential for a parent nucleus exhibiting binary fission is given by,

$$V = \frac{Z_1 Z_2 e^2}{r} + V_p(z) + \frac{\hbar^2 \ell(\ell+1)}{2\mu r^2} \quad , \text{ for } z > 0. \qquad (2)$$

Here $Z_1$ and $Z_2$ are the atomic numbers of the binary fission fragments, '$z$' is the distance between the near surfaces of the fragments, '$r$' is the distance between fragment centers and is given as $r = z + C_1 + C_2$, where, $C_1$ and $C_2$ are the Süsmann central radii of fragments. The term $\ell$ represents the angular momentum, $\mu$ the reduced mass and $V_P$ is the proximity potential. The proximity potential $V_P$ is given by Blocki et al., [48,49] as,

$$V_p(z) = 4\pi \gamma b \left[ \frac{C_1 C_2}{(C_1 + C_2)} \right] \Phi\left(\frac{z}{b}\right), \qquad (3)$$

with the nuclear surface tension coefficient,
$$\gamma = 0.9517[1 - 1.7826(N - Z)^2 / A^2] \text{ MeV/fm}^2, \qquad (4)$$
where $N$, $Z$ and $A$ represent neutron, proton and mass number of parent respectively, $\Phi$ represents the universal proximity potential [49] given as,
$$\Phi(\varepsilon) = -4.41 e^{-\varepsilon/0.7176}, \text{ for } \varepsilon > 1.947, \qquad (5)$$
$$\Phi(\varepsilon) = -1.7817 + 0.9270\varepsilon + 0.0169\varepsilon^2 - 0.05148\varepsilon^3, \text{ for } 0 \leq \varepsilon \leq 1.9475 \qquad (6)$$
with $\varepsilon = z/b$, where the width (diffuseness) of the nuclear surface $b \approx 1$ fm and Süsmann central radii $C_i$ of fragments related to sharp radii $R_i$ as,

$$C_i = R_i - \left(\frac{b^2}{R_i}\right) \qquad (7)$$

For $R_i$ we use semi empirical formula in terms of mass number $A_i$ as [48],
$$R_i = 1.28 A_i^{1/3} - 0.76 + 0.8 A_i^{-1/3} \qquad (8)$$
The potential for the internal part (overlap region) of the barrier is given as,
$$V = a_0 (L - L_0)^n, \text{ for } z < 0 \qquad (9)$$

Here $L = z + 2C_1 + 2C_2$ and $L_0 = 2C$, the diameter of the parent nuclei. The constants $a_0$ and $n$ are determined by the smooth matching of the two potentials at the touching point.

Using one-dimensional WKB approximation, the barrier penetrability P is given as,

$$P = \exp\left\{-\frac{2}{\hbar}\int_a^b \sqrt{2\mu(V-Q)}\,dz\right\} \quad (10)$$

Here the mass parameter is replaced by $\mu = mA_1A_2/A$, where 'm' is the nucleon mass and $A_1$, $A_2$ are the mass numbers of binary fission fragments respectively. The turning points 'a' and 'b' are determined from the equation $V(a) = V(b) = Q$.

The relative yield can be calculated as the ratio between the penetration probability of a given fragmentation over the sum of penetration probabilities of all possible fragmentation as follows,

$$Y(A_i, Z_i) = \frac{P(A_i, Z_i)}{\sum P(A_i, Z_i)} \quad (11)$$

### 3. Results, discussion and conclusion

Using the concept of cold reaction valley the binary fission of even-even $^{244-258}$Cf isotopes has been studied. In the study, the structure of minima in the driving potential is considered. The driving potential is defined as the difference between the interaction potential, $V$ and the decay energy, $Q$ of the reaction. Most of the $Q$ values are calculated using experimental mass excesses of Audi et al., [50] and some masses are taken from the table of KUTY [51]. The interaction potential is calculated as the sum of Coulomb and proximity potentials. Next the driving potential ($V-Q$) for a particular parent nuclei is calculated for all possible fission fragments as a function of mass and charge asymmetries respectively given as $\eta = \dfrac{A_1 - A_2}{A_1 + A_2}$ and $\eta_Z = \dfrac{Z_1 - Z_2}{Z_1 + Z_2}$, at the touching configuration. For every fixed mass pair ($A_1$, $A_2$) a pair of charges is singled out for which the driving potential is minimized.

### 3.1 Cold reaction valley of even – even $^{244-258}$Cf isotopes

The driving potential for the touching configuration of fragments are calculated for the binary fragmentation of even-even $^{244-258}$Cf isotopes as the representative parent nucleus. Fig. 1-4 represent the plots for driving potential versus $A_1$ (mass of one fragment) for even - even $^{244-258}$Cf isotopes respectively. The occurrences of the mass-asymmetry valleys in these figures are due to the shell effects of one or both the fragments. The fragment combinations corresponding to the minima in the potential energy will be the most probable binary fission fragments.

From Fig. 1-4 we found that the first minimum in each plot corresponds to the splitting $^4$He+$^{240}$Cm, $^4$He+$^{242}$Cm, $^4$He+$^{244}$Cm, $^4$He+$^{246}$Cm, $^4$He+$^{248}$Cm, $^4$He+$^{250}$Cm, $^4$He+$^{252}$Cm and $^4$He+$^{254}$Cm for even-even $^{244-258}$Cf isotopes respectively and these fragment combination shows the deepest minimum in the cold valley. For $^{244}$Cf in addition to the alpha particle $^{8,10}$Be, $^{14}$C, $^{34}$Si, $^{38,40}$S, $^{44}$Ar, $^{48,50}$Ca, $^{80}$Ge, $^{84}$Se, $^{88}$Kr etc. are found to be the possible candidates for emission. Moving on to the fission region, there are three deep regions each consisting of few minima. For the first valley as one can see from Fig. 1 (a), the first minimum corresponds to the splitting $^{34}$Si+$^{210}$Po, while the second and third minima correspond to the splitting $^{38}$S+$^{206}$Pb and $^{40}$S+$^{204}$Pb. From the cold valley approach the first minimum is due to the magic neutron shell N

= 126 of $^{210}$Po and magic neutron shell N = 20 of $^{34}$Si, the second and third minimum is occurring due to the magic proton shell Z = 82 of $^{206}$Pb and $^{204}$Pb respectively. Other fragment combinations in this region are $^{48}$Ca+$^{196}$Pt and $^{50}$Ca+$^{194}$Pt, due to the presence of doubly magic $^{48}$Ca (N = 28 and Z = 20) and proton shell closure Z = 20 of $^{50}$Ca. In second valley the splitting $^{84}$Se+$^{160}$Gd is due to the presence of magic shell N = 50 of $^{84}$Se. In the case of the third valley, the first two minima involve $^{108}$Ru+$^{136}$Xe and $^{110}$Pd+$^{134}$Te splitting and therefore their occurrence is attributed to the presence of magic neutron shell N = 82 of $^{136}$Xe and $^{134}$Te. Other minima in this valley comes from the splitting $^{116}$Cd+$^{128}$Sn, $^{118}$Cd+$^{126}$Sn, $^{120}$Cd+$^{124}$Sn and $^{122}$Cd+$^{122}$Sn, due to the presence of Z = 50 magic shell.

    Just as in the case of $^{244}$Cf, even-even $^{246-258}$Cf isotopes also has three deep valleys in the fission regions each consisting of several comparable minima. In the first region the minima obtained for $^{246}$Cf isotope is at doubly magic $^{206,208}$Pb, $^{200,202}$Hg and $^{48}$Ca. For $^{248,250}$Cf isotope, the minima is obtained for near doubly magic $^{204}$Hg and doubly magic $^{208}$Pb whereas for $^{252,254}$Cf isotope the minima is obtained for $^{206}$Hg and $^{50,52}$Ca possessing magic shell N = 126 and Z = 20 respectively. The minima for $^{256}$Cf isotope are at $^{208}$Hg and $^{52}$Ca whereas for $^{258}$Cf isotope the minima are at $^{210}$Hg and $^{52,54}$Ca. In the second region the minima at $^{82}$Ge and $^{84}$Se due to magic shell N = 50 are found for $^{246,248,250}$Cf isotope. For $^{252,254,256}$Cf isotope, the minimum is found for $^{82}$Ge whereas minimum at $^{80}$Zn is obtained for $^{254,256,258}$Cf isotopes. The minimum around Ni due to magic shell Z = 28 is obtained for $^{252,258}$Cf isotope. Finally, in the third valley the minimum at $^{132}$Te is found for $^{246,248}$Cf isotope whereas a nearly doubly magic nucleus $^{134}$Te is obtained for $^{246,248,250,252}$Cf isotope. Also in this region the minima is obtained around $^{126,128}$Sn due to magic shell Z=50 for $^{246}$Cf and around $^{132}$Sn for $^{250,252,254,256,258}$Cf isotope.

    It is clear from Fig. 1-4 that, as we move towards the symmetric fission region, we can see that the driving potential decreases with increase in mass number (ie. due to the increase in neutron number) of the parent nuclei. This is because in this region there is a chance for symmetric fission to occur (for e.g. $^{124}$Sn + $^{124}$Cd, $^{130}$Sn + $^{128}$Cd). This also stresses the role of double or near double magicity of the fragments. It is evident from Fig. 4 (a) that in the case of $^{256}$Cf isotope the minimum observed at $^{124}$Cd+$^{132}$Sn is almost near to the deepest minimum found at $^{4}$He+$^{252}$Cm whereas in the case of $^{258}$Cf isotope it is clear from Fig. 4 (b) that the minimum found at $^{126}$Cd+$^{132}$Sn is comparable with that obtained at $^{4}$He+$^{254}$Cm.

## 3.2 Barrier penetrability and Yield calculation

    The barrier penetrability for each charge minimized fragment combinations found in the cold valley for even – even $^{244-258}$Cf isotopes are calculated using the formalism described above. Using eqn. (10) relative yield is calculated and is plotted as a function of fragment mass number $A_1$ and $A_2$ in Fig. 5–8. The most favorable fragment combinations for all the eight isotopes mentioned above are obtained by calculating their relative yield.

    From Fig. 5(a), it is clear that for $^{244}$Cf, the combination $^{36}$S+$^{208}$Pb possesses highest yield due to the presence of doubly magic nuclei $^{208}$Pb (N = 126, Z = 82). The next higher yield can be observed for the $^{34}$Si+$^{210}$Po combination and is due to the near doubly magic $^{210}$Po (N = 126, Z = 84). The various other fragment combinations observed in this binary fission of parent nuclei $^{244}$Cf are $^{68}$Ni+$^{176}$Yb, $^{70}$Ni+$^{174}$Yb, $^{108}$Ru+$^{136}$Xe, $^{110}$Pd+$^{134}$Te. Of these the first and second one are attributed to the magic shell Z = 28 of Ni while the third fragment combination is due to the presence of neutron shell closure at N = 82 of $^{136}$Xe. The fragment combination with $^{134}$Te is due to the near double magicity Z = 52 and N = 82. The splitting $^{116}$Cd+$^{128}$Sn, $^{118}$Cd+$^{126}$Sn and $^{120}$Cd+$^{124}$Sn are due to the presence of magic number Z = 50 of Sn.

In the case of $^{246}$Cf isotope, $^{38}$S+$^{208}$Pb is the most favored binary splitting and it is due to the presence of doubly magic $^{208}$Pb (Z = 82, N = 126). The next higher yield is observed for the $^{40}$S+$^{206}$Pb combination and is due to the near doubly magic $^{206}$Pb (N = 124, Z= 82). The various fragment combinations found in the binary fission process are $^{34}$Si+$^{212}$Po, $^{46}$Ar+$^{200}$Hg, $^{48}$Ca+$^{198}$Pt, $^{110}$Ru+$^{136}$Xe, $^{112}$Pd+$^{134}$Te, $^{114}$Pd+$^{132}$Te $^{118}$Cd+$^{128}$Sn, $^{120}$Cd+$^{126}$Sn and $^{122}$Cd+$^{124}$Sn. The first combination is due to the near doubly magic $^{212}$Po (N = 128, Z = 84). The second combination is due to neutron shell closure N= 28 of $^{46}$Ar and also due to near proton shell closure Z= 80 of $^{200}$Hg. The third combination is due to doubly magic $^{48}$Ca (Z = 20, N = 28). The fourth combination is due to neutron shell closure N=82 of $^{136}$Xe. The fifth and sixth combinations are due to the near doubly magic $^{134}$Te (N = 82, Z = 52) and $^{132}$Te (N = 80, Z = 52) respectively. The last three combinations are attributed to the magic shell closure at Z = 50 of $^{128}$Sn, $^{126}$Sn and $^{124}$Sn respectively.

For $^{248}$Cf isotope, the highest yield is obtained for the fragment combination $^{40}$S+$^{208}$Pb due to the presence of doubly magic $^{208}$Pb (N = 126, Z = 82). The next higher yields are for the fragment combinations $^{46}$Ar+$^{202}$Hg, $^{44}$Ar+$^{204}$Hg and $^{48}$Ca+$^{200}$Pt which possess the neutron shell closure N =28 of $^{46}$Ar, near proton shell closure Z = 80 of $^{202}$Hg, near doubly magic shell of $^{204}$Hg (N = 124, Z = 80) and the doubly magic $^{48}$Ca (N = 28, Z = 20). The various fragment combinations that occur in this binary fission process are $^{34}$Si+$^{214}$Po, $^{70}$Ni+$^{178}$Yb, $^{72}$Ni+$^{176}$Yb and $^{78}$Zn+$^{170}$Er. These splittings are due to the neutron shell closure at N = 20 of $^{34}$Si, near proton shell closure at Z = 84 of $^{214}$Po, proton shell closure at Z = 28 of Ni and near neutron shell closure N = 48 of $^{78}$Zn respectively.

In the case of $^{250}$Cf isotope, the highest yield is obtained for the fragment combination $^{46}$Ar+$^{204}$Hg due to the presence of nearly doubly magic $^{204}$Hg (N = 124, Z = 80). The next higher yields are for the fragment combinations $^{42}$S+$^{208}$Pb, $^{44}$Ar+$^{206}$Hg, $^{40}$S+$^{210}$Pb, $^{48}$Ca+$^{202}$Pt and $^{50}$Ca+$^{200}$Pt. It is due to the doubly magic $^{208}$Pb (N = 126, Z = 82), magic shell N=126 of $^{206}$Hg, magic shell Z=82 of $^{210}$Pb, doubly magic $^{48}$Ca (N = 28, Z = 20), magic shell Z=20 of $^{50}$Ca.

For $^{252}$Cf isotope, the highest yield is obtained for the fragment combination $^{46}$Ar+$^{206}$Hg due to the presence of nearly doubly magic $^{206}$Hg (N = 126, Z = 80). The next higher yields are for the fragment combinations $^{122}$Cd+$^{130}$Sn, $^{120}$Cd+$^{132}$Sn, $^{50}$Ca+$^{202}$Pt, $^{124}$Cd+$^{128}$Sn and $^{42}$S+$^{210}$Pb. It is due to the magic shell Z = 50 of $^{130}$Sn, doubly magic $^{132}$Sn (N = 82, Z = 50), magic shell Z=20 of $^{50}$Ca, Z=50 of $^{128}$Sn, Z=82 of $^{210}$Pb.

In the case of $^{254}$Cf isotope, the highest yield is obtained for the fragment combination $^{122}$Cd+$^{132}$Sn due to the presence of doubly magic $^{132}$Sn (N = 82, Z = 50). The next higher yields are for the fragment combinations $^{124}$Cd+$^{130}$Sn, $^{126}$Cd+$^{128}$Sn, $^{48}$Ar+$^{206}$Hg, $^{50}$Ca+$^{204}$Pt, $^{52}$Ca+$^{202}$Pt and $^{46}$Ar+$^{208}$Hg. The first two combinations are due to magic shell Z = 50 of $^{130}$Sn and $^{128}$Sn whereas third one is due to neutron shell closure N = 126 and near proton shell closure Z = 80 of $^{206}$Hg. The combination $^{50}$Ca+$^{204}$Pt is due to proton shell closure Z= 20 of $^{50}$Ca and neutron shell closure N= 126 of $^{204}$Pt. The splitting $^{52}$Ca+$^{202}$Pt is due to proton shell closure Z = 20 of $^{52}$Ca and near neutron shell closure N= 124 of $^{202}$Pt. The last combination is due to near doubly magic $^{208}$Hg (N = 128, Z =80).

For $^{256}$Cf isotope, the highest yield is obtained for the fragment combination $^{124}$Cd+$^{132}$Sn due to the presence of doubly magic $^{132}$Sn (N = 82, Z = 50). The next higher yields are for the fragment combinations $^{126}$Cd+$^{130}$Sn, $^{128}$Cd+$^{128}$Sn, $^{52}$Ca+$^{204}$Pt, $^{122}$Pd+$^{134}$Te, $^{48}$Ar+$^{208}$Hg and $^{46}$Ar+$^{210}$Hg. The first two combinations are due to magic shell Z = 50 of $^{130}$Sn and $^{128}$Sn. The splitting $^{52}$Ca+$^{204}$Pt and $^{122}$Pd+$^{134}$Te are due to proton shell closure Z = 20 of $^{52}$Ca, neutron shell closure N= 126 of $^{204}$Pt and near doubly magic $^{134}$Te (N = 82, Z =52). The last combinations are due to near proton shell closure (Z=80) of $^{208,210}$Hg.

In the case of $^{258}$Cf isotope, the fragment combination $^{126}$Cd+$^{132}$Sn possesses the highest yield due to the presence of doubly magic $^{132}$Sn (N = 82, Z = 50). The next higher yields are for the fragment combinations $^{128}$Cd+$^{130}$Sn, $^{124}$Cd+$^{134}$Sn, $^{52}$Ca+$^{206}$Pt, $^{54}$Ca+$^{204}$Pt, $^{48}$Ar+$^{210}$Hg and $^{46}$Ar+$^{212}$Hg. The first two combinations are due to magic shell Z = 50 of $^{130}$Sn and $^{134}$Sn. The combination $^{52}$Ca+$^{206}$Pt is due to proton shell closure Z= 20 of $^{52}$Ca and near neutron shell closure N= 128 of $^{206}$Pt. The splitting $^{54}$Ca+$^{204}$Pt is due to proton shell closure Z= 20 of $^{54}$Ca and neutron shell closure N= 126 of $^{204}$Pt. The last two combinations are due to near proton shell closure Z = 80 of $^{210,212}$Hg and neutron shell closure N = 28 of $^{46}$Ar.

Our work reveals that, the presence of doubly magic or near doubly magic nuclei plays an important role in the binary fission of even-even $^{244-258}$Cf isotopes. It is found that the magnitude of the relative yield increases with increase in mass number (ie. due to the increase in neutron number) of the parent nuclei. Also it is found that highest yield for $^{244,246,248}$Cf isotopes are for the fragments with isotope of Pb (Z=82) as one fragment, whereas for $^{250}$Cf and $^{252}$Cf isotopes the highest yield is for the fragments with isotope of Hg (Z=80) as one fragment. The highest yield (or minima in the cold reaction valley) associated with $^{208}$Pb daughter can be interpreted as a heavy particle radioactivity [52-55]. In the case of $^{254,256,258}$Cf isotopes the highest yield is for the fragments with Sn (Z=50) as one fragment. It is found that for the binary fragmentation of $^{244,246,248,250}$Cf isotopes, asymmetric splitting is dominant and in the case of $^{254,256,258}$Cf isotopes symmetric splitting is dominant. In the case of $^{252}$Cf isotope, the highest yield is for the fragment combination $^{46}$Ar+$^{206}$Hg, which corresponds to asymmetric splitting whereas the second highest yield is obtained for the fragmentation $^{122}$Cd+$^{130}$Sn that corresponds to symmetric splitting. Hence, we can say that both asymmetric splitting and symmetric splitting are favorable for the binary fission of $^{252}$Cf isotope. In Fig. 9, we have compared the individual yields obtained for the cold fission of $^{252}$Cf isotope with the experimental data taken from the γ- γ- γ coincidences technique using Gammasphere [12,14].

## 4. Summary

The binary fragmentation of even-even $^{244-258}$Cf isotopes has been studied by taking Coulomb and proximity potential as interacting barrier. In each case, the fragmentation potential and Q-values are calculated for all possible fission components, which reveal that, the even mass number fragments are more favored than odd mass number fragments. The favorable fragment combinations are obtained by calculating the relative yield. It is found that highest yield for $^{244,246,248}$Cf isotopes are for the fragments with isotope of Pb ( Z=82) as one fragment, whereas for $^{250}$Cf and $^{252}$Cf isotopes the highest yield is for the fragments with isotope of Hg (Z=80) as one fragment. In the case of $^{254,256,258}$Cf isotopes, the highest yield are for the fragments with Sn (Z=50) as one fragment. This reveals the role of doubly magic and near doubly magic shell closures (of $^{48}$Ca, $^{132}$Sn, $^{134}$Te, $^{204}$Hg and $^{208}$Pb) in binary fission.

## Acknowledgments

One of the author (KPS) would like to thank the University Grants Commission, Govt. of India for the financial support under Major Research Project. No.42-760/2013 (SR) dated 22-03-2013.

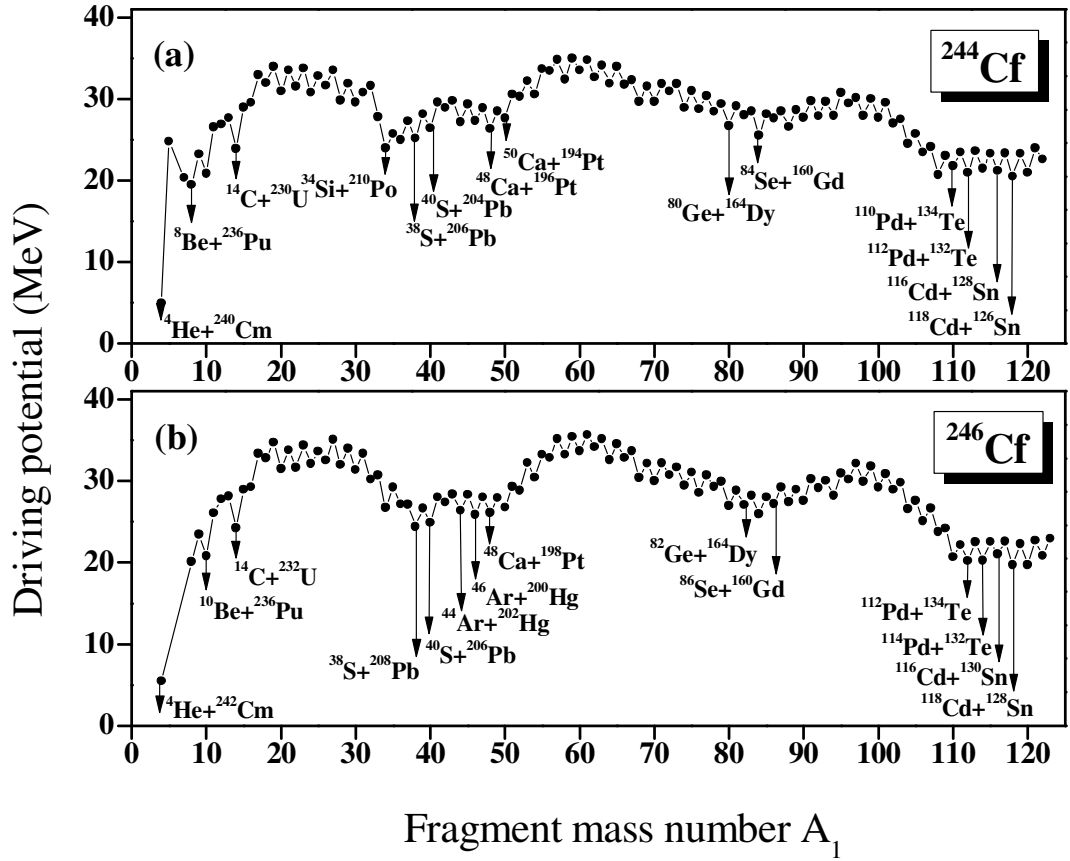

**Fig.1.** The driving potential for $^{244}$Cf and $^{246}$Cf isotope plotted as a function of mass number $A_1$.

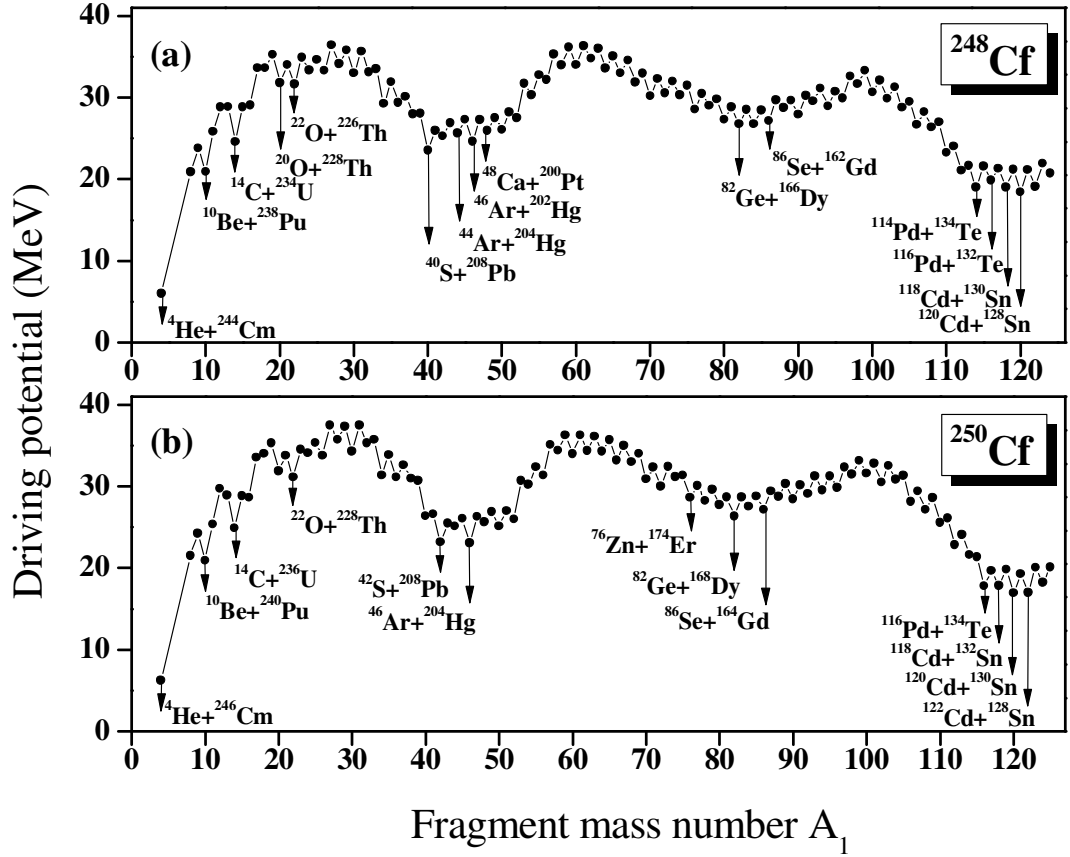

**Fig.2.** The driving potential for $^{248}$Cf and $^{250}$Cf isotope plotted as a function of mass number $A_1$.

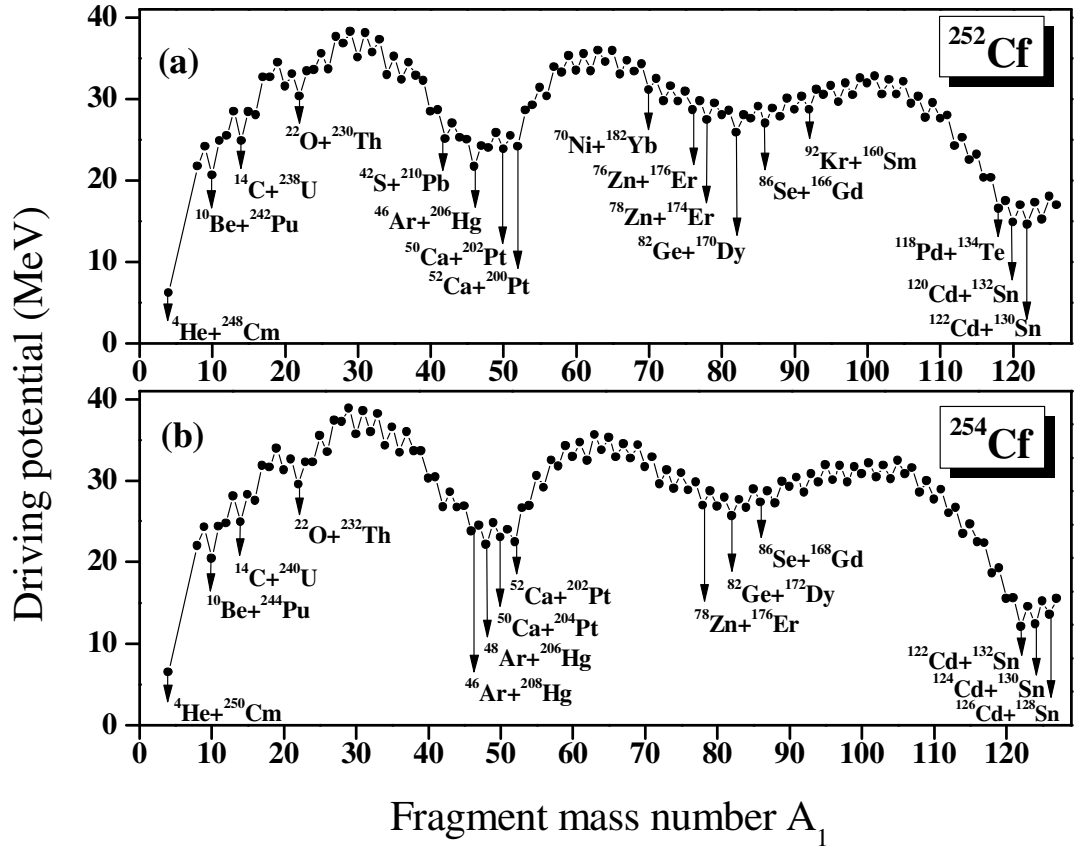

**Fig. 3.** The driving potential for $^{252}$Cf and $^{254}$Cf isotope plotted as a function of mass number $A_1$.

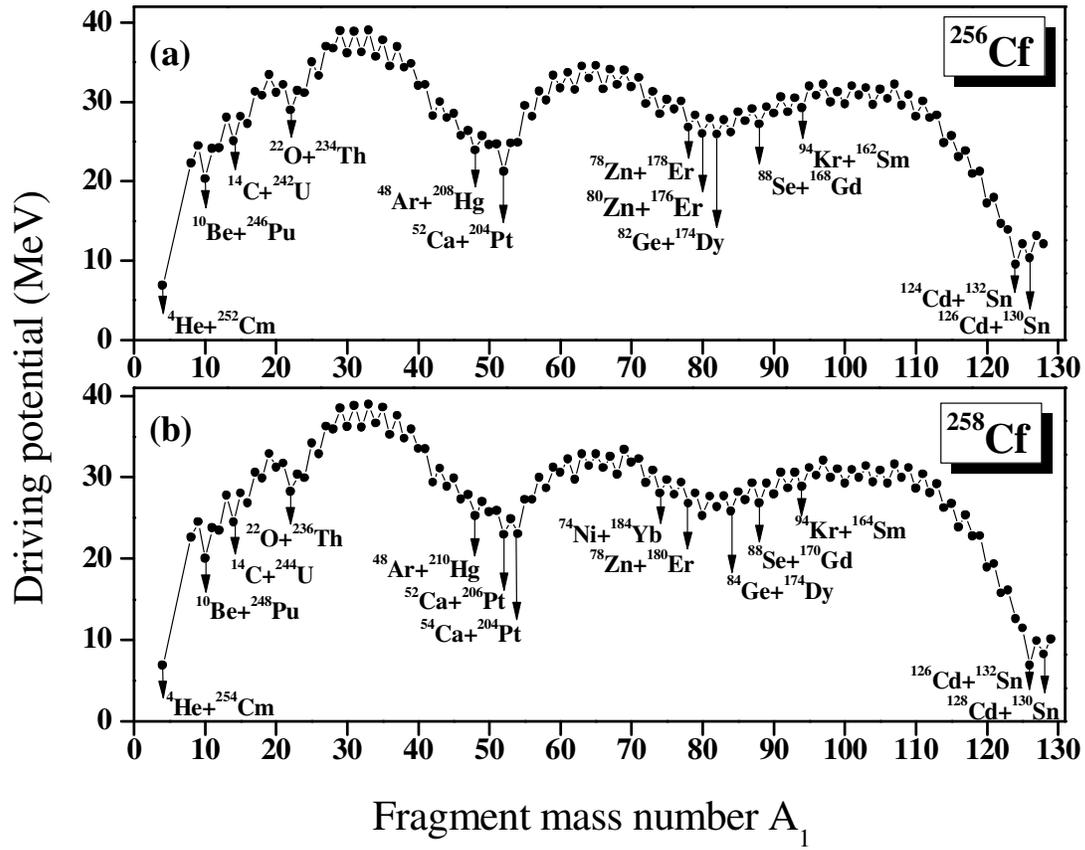

**Fig. 4.** The driving potential for $^{256}$Cf and $^{258}$Cf isotope plotted as a function of mass number $A_1$.

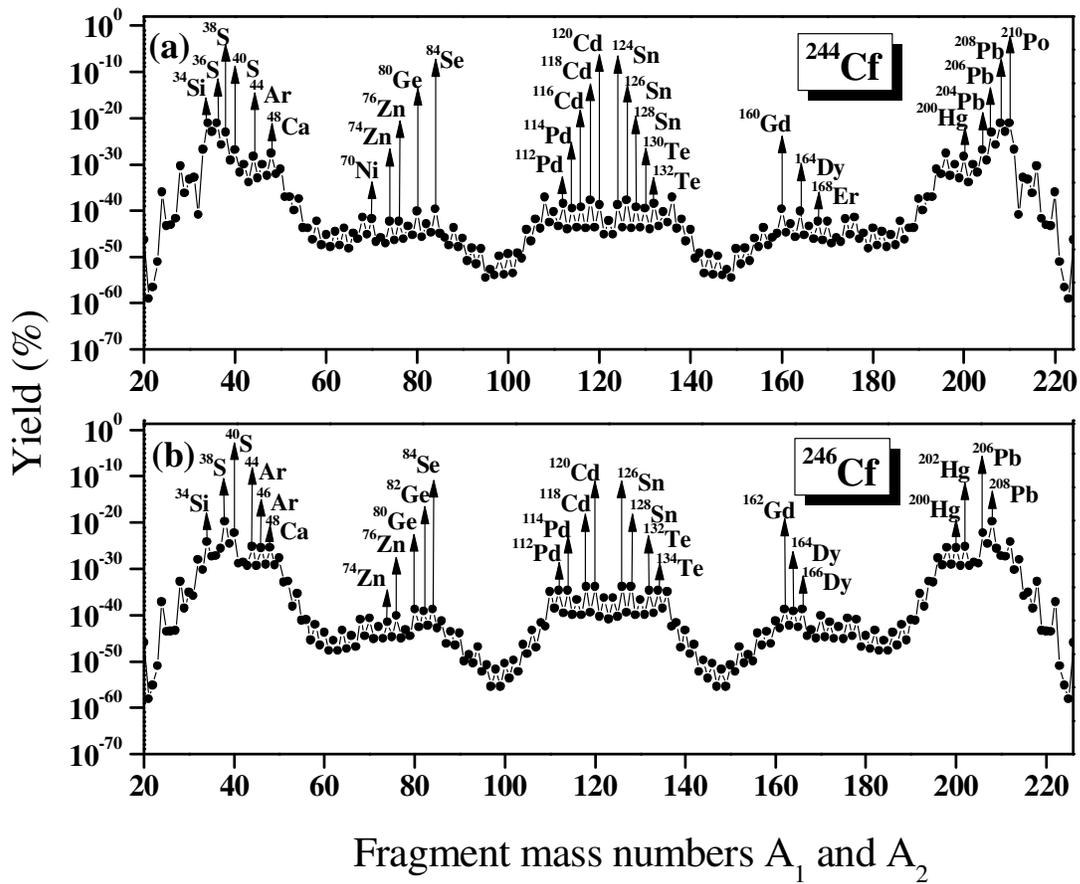

**Fig. 5.** The relative yields plotted as a function of mass numbers $A_1$ and $A_2$ for $^{244}$Cf and $^{246}$Cf isotope. The fragment combinations with higher yields are labeled.

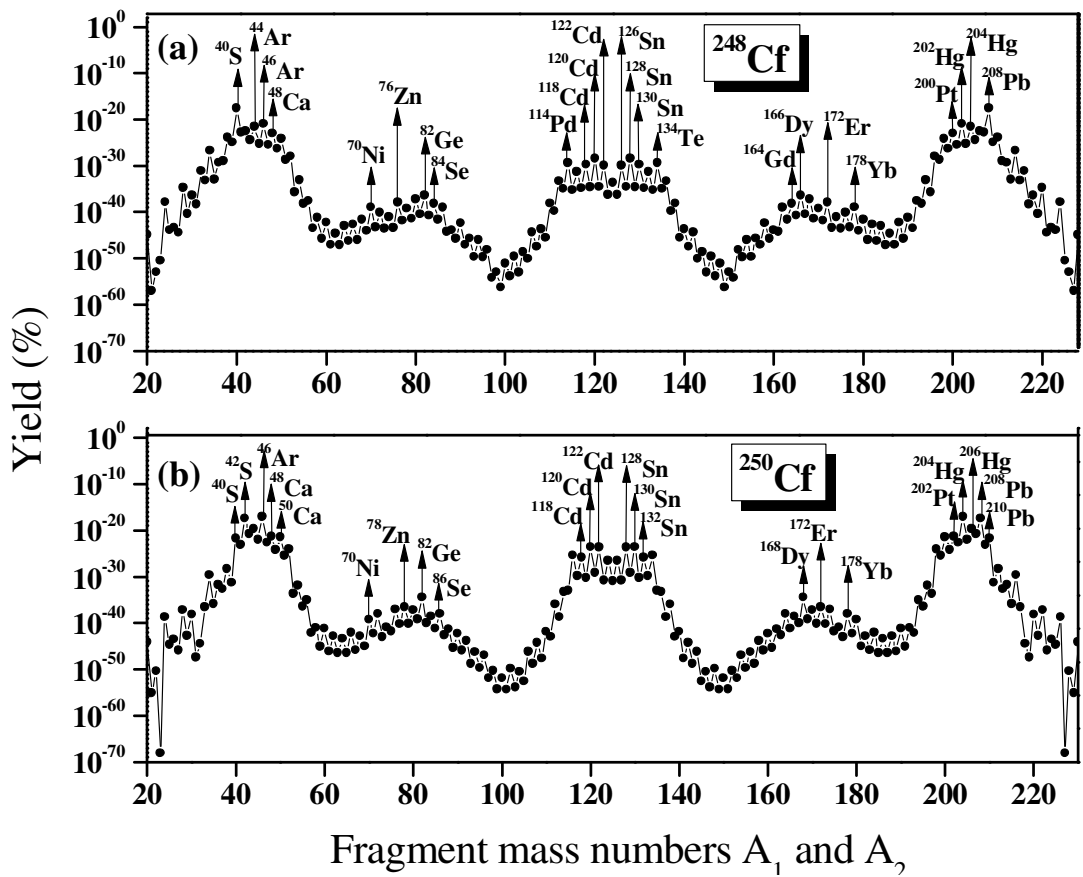

**Fig. 6.** The relative yields plotted as a function of mass numbers $A_1$ and $A_2$ for $^{248}$Cf and $^{250}$Cf isotope. The fragment combinations with higher yields are labeled.

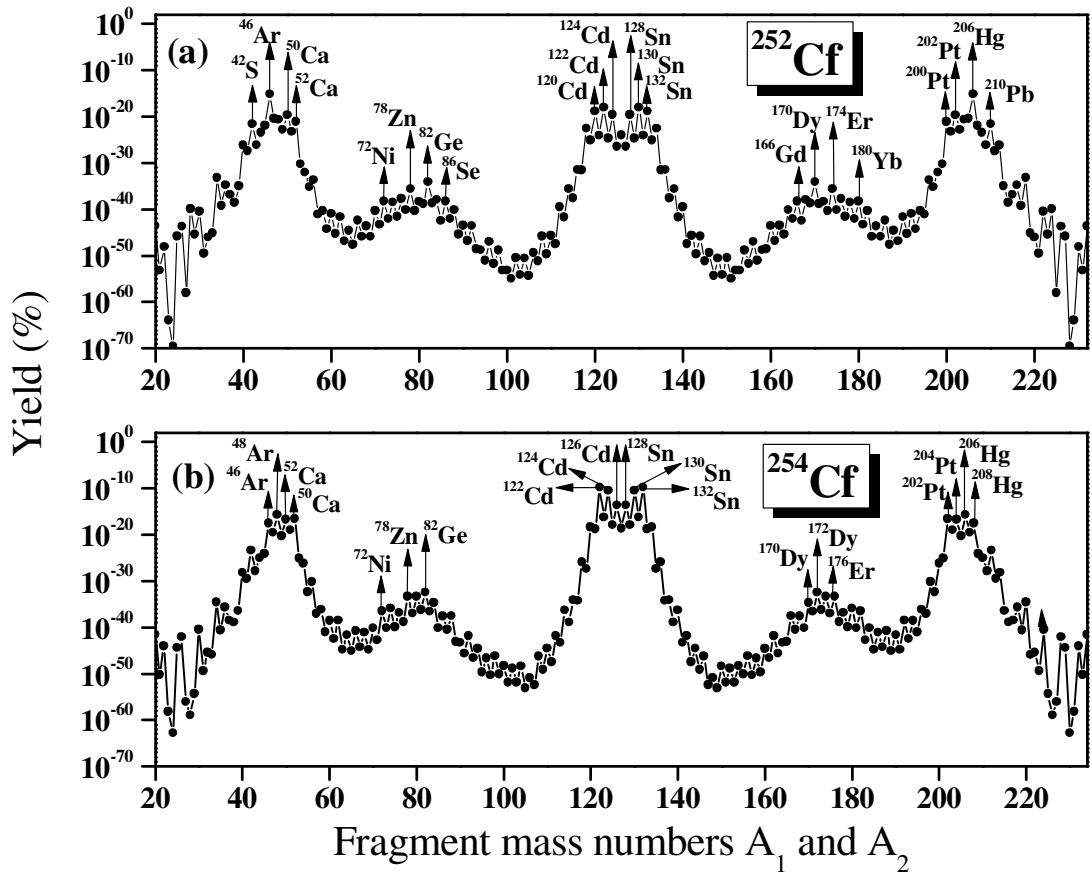

**Fig. 7.** The relative yields plotted as a function of mass numbers $A_1$ and $A_2$ for $^{252}$Cf and $^{254}$Cf isotope. The fragment combinations with higher yields are labeled.

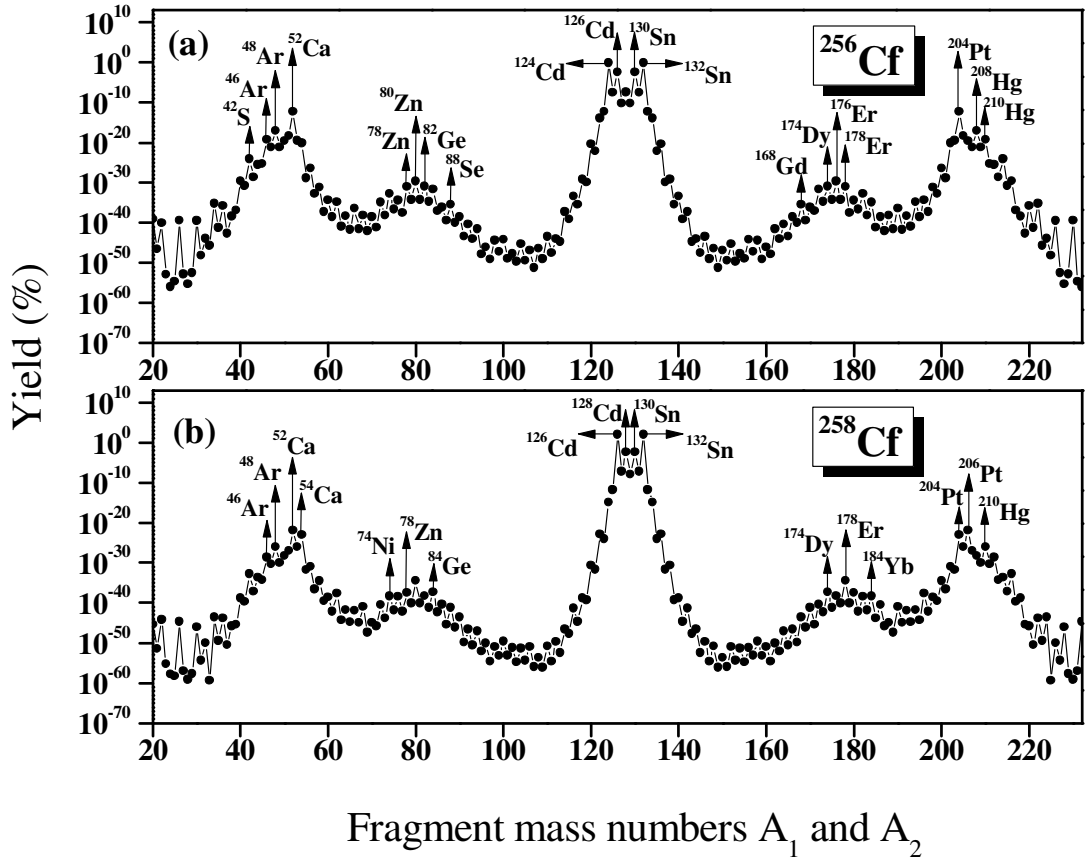

Fig. 8. The relative yields plotted as a function of mass numbers $A_1$ and $A_2$ for $^{256}$Cf and $^{258}$Cf isotope. The fragment combinations with higher yields are labeled.

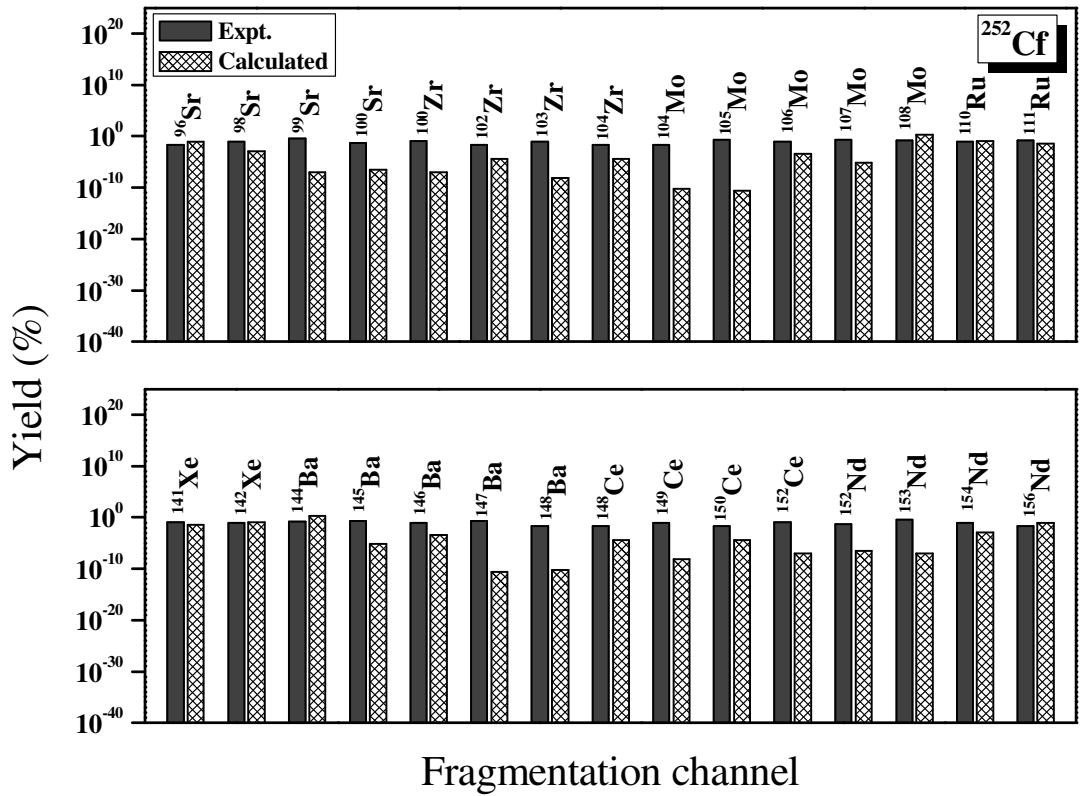

**Fig.9.** The yields obtained for the cold fission of $^{252}$Cf isotope and their comparison with the experimental data [12,14].